# Holographic lasing with dielectric metasurfaces

Ayesheh Bashiri[1,2]*, Aleksandr Vaskin[2], Muyi Yang[1,2,4], Katsuya Tanaka[1,2,4], Marijn Rikers[1,2,5], Thomas Pertsch[2,3,4], Isabelle Staude[1,2,4]

[1]Institute of Solid-State Physics, Friedrich Schiller University; Jena, 07743, Germany.
[2]Abbe Center of Photonics, Institute of Applied Physics, Friedrich Schiller University; Jena, 07745, Germany.
[3]Fraunhofer-Institute of Applied Optics and Precision Engineering IOF; Jena, 07745, Germany.
[4]Max Planck School of Photonics; Jena, 07745, Germany.
[5]Department of Quantum Science and Technology, Research School of Physics, Australian National University; Canberra, ACT2601, Australia.

*Corresponding author. Email: Ayesheh.bashiri@uni-jena.de

**Abstract:**

Light-emitting metasurfaces provide a compact, integrated solution for simultaneous light generation and beam shaping, making them a promising candidate for advanced photonic applications. However, existing approaches for tailoring far-field emission patterns primarily operate in the spontaneous emission regime, where low coherence limits precise light control. Here, we present a light-emitting metasurface system with holographic lasing output, composed of a binary-structured metasurface integrated with a gain medium, that enables coherent light generation and precise beam shaping of the output lasing emission. With a compact footprint, low lasing threshold, and wide field of view, our system offers an exceptional platform for generating arbitrarily-structured lasing light, with significant potential for miniaturized optical systems.

## Introduction:

Developing an ultra-compact laser source with built-in spatial wavefront shaping remains a major challenge in nanophotonics, with significant implications for augmented reality (AR), light field displays, deep diffractive neural networks, and next-generation medical diagnostics and treatment. Metasurfaces, planar optical elements enabling wavefront control, have emerged as promising tools in this context (*1-4*). When combined with active materials (*5,6*), light-emitting metasurfaces can simultaneously generate and shape light (*7*); however, operating in the spontaneous emission regime, they suffer from limited temporal and spatial coherence, restricting their ability to form interference-based complex patterns.
In contrast, operating in the lasing regime establishes coherence among emitted photons, producing light with well-defined phase relationships and enabling precise wavefront shaping for holographic applications. Lasing metasurfaces, which integrate resonant nanostructures with optical gain, have recently emerged as compact, coherent light sources (*8-11*). However, most remain restricted to simple, fixed-direction beams, lacking the capability to project complex spatial patterns.

One approach to overcome this limitation is to integrate metasurface reflectors into external cavity lasers (ECL) as the beam-shaping element (*12*). However, they remain limited by reliance on external laser diodes, precise alignment, and bulky configurations. Another strategy employs metasurface-integrated VCSELs, with dielectric metasurfaces monolithically fabricated atop VCSEL apertures for directional beam shaping (*13,14*). While offering better integration, these devices face challenges such as complex epitaxial growth and cavity thickness exceeding tens of microns. Moreover, placing the metasurface outside the cavity limits its function to post-emission beam shaping. In contrast, embedding the metasurface within the laser cavity additionally allows for direct control over lasing threshold, wavelength, and polarization. This approach enhances light–matter interaction and local density of optical states (LDOS), boosting the Purcell factor and spontaneous emission coupling ratio, enabling low-threshold, multifunctional lasing in compact, subwavelength architectures. Yet, simultaneous generation and arbitrary shaping of laser emission within a single metasurface remains largely unexplored.

Here, we report the first demonstration of a binary holographic metasurface laser that unites coherent light generation and pixel-level spatial control in a single compact device (Fig. 1). The system consists of a titanium dioxide ($TiO_2$) dielectric metasurface, spin-coated with a SU8 resist layer doped with Rhodamine 6G laser dye as the gain medium. The latter is chosen for its high quantum yield, broad gain spectrum, and ease of integration. Our design employs a binary "on–off" metasurface configuration, wherein spatial modulation of the emission is encoded by the presence (on) or absence (off) of nanoresonators across the array. This enables arbitrary holographic pattern projection via engineered intracavity feedback. With a compact footprint of around 37 μm×37 μm and submicron thickness, a relatively low lasing threshold of less than 8 nJ (pump energy per pulse), and the ability to project patterns over a high FOV of 100°×100°, the device demonstrates exceptional performance for generating structured light. As a proof of concept, we have successfully projected the abbreviation "FSU" in the holographic output, representing "Friedrich-Schiller-University". This work bridges the gap between passive metasurfaces and conventional laser systems, introducing a new class of coherent light sources that seamlessly integrate holography and lasing within a single, ultra-flat platform. This paves the way for structured-light applications, such as AR contact lenses, where each device functions as both light emitter and projector, for nonlinear hidden layers of diffractive neural networks exploiting the inherent optical nonlinearities associated with lasing action, or for light field displays, where individual metasurface lasers could serve as scalable, self-contained pixel-microlens units.

Our approach relies on second-order distributed feedback (DFB) and guided-mode resonance to achieve low-threshold lasing. The periodic high-index nanoresonator array forms a photonic bandgap for laterally guided modes in the SU8 layer by fulfilling the second-order Bragg diffraction condition at the design wavelength, thereby enabling in-plane optical feedback essential for lasing (*8*). In addition to coupling to guided modes, the resonant scatterers support collective surface-lattice resonance (SLR)-like interactions (*15*), further enhancing feedback through diffractive coupling. Near the photonic band edge, the dispersion flattens, leading to a slow-light regime with substantially enhanced LDOS. This enhances light–matter interaction within the gain medium and significantly raises the Q-factor by extending photon lifetimes and suppressing radiative losses. Using low-loss dielectric materials minimizes dissipation, further enhancing the Q-factor and feedback efficiency. Second-order Bragg diffraction additionally enables vertical outcoupling of the guided modes, supporting lasing emission normal to the surface. However, by spatially patterning the metasurface—through the binary presence or absence of nanoresonators—this emission is redistributed into a broad angular range, allowing structured light projection over a wide field of view.

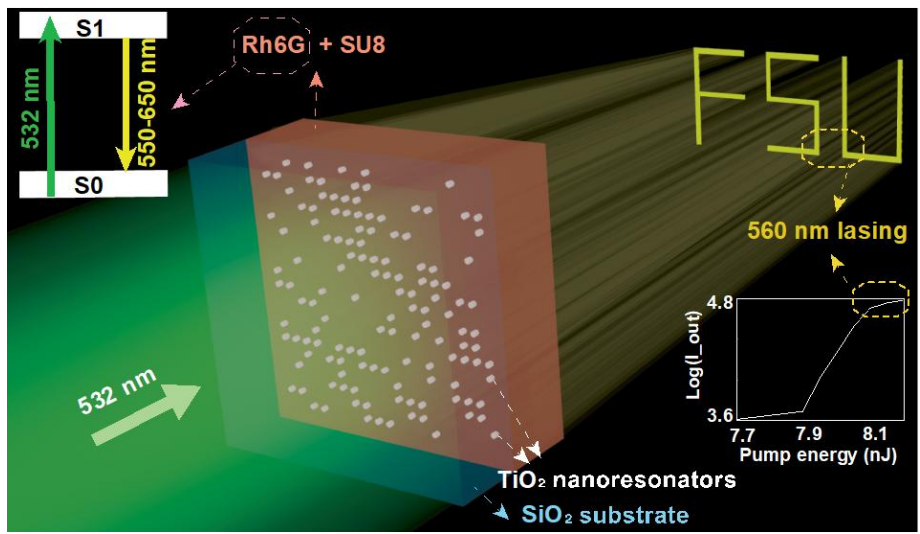

**Fig. 1. Schematic illustration of on-off binary lasing metasurface.** An engineered array of titanium dioxide ($TiO_2$) nanoresonators on a silicon dioxide ($SiO_2$) substrate is integrated with an SU8 layer incorporated with Rhodamine 6G (Rh6G) laser dye. Upon a 532 nm excitation, the coupled system generates lasing light at 560 nm with an arbitrary emission pattern in the far-field.

**Results and discussion:**

The design of our lasing metasurface was performed in two subsequent steps. First, using the finite element method (FEM) implemented by the software package COMSOL Multiphysics, we designed a periodic metasurface (*16*) composed of a 2-dimensional (2D) square array of resonant TiO$_2$ (see fig. S1 for refractive indices) nanocylinders on a SiO$_2$ substrate and an SU8 photoresist coating acting as a 2D planar waveguide. The optical confinement along the out-of-plane direction leads to a strong in-plane amplification of spontaneous emission in favor of lasing. Note that during the experimental stage, rhodamine 6G (Rh6G) laser dye was integrated into the SU8 film to form the gain medium.

In the design process, we optimized the nanocylinders dimensions, array period, and SU8 film thickness to achieve three key objectives: 1) high Q-factor resonances excited under normal incidence and overlapping with the Rh6G emission spectrum (Fig. 2A), 2) strong field enhancement within the gain medium, and 3) the formation of a band structure with a wide stop-band at the Γ point to enhance the DFB mechanism for out-of-plane lasing (*8*). The array period is chosen to satisfy the second-order Bragg diffraction condition, providing in-plane optical feedback. Simultaneously, the nanocylinders are designed to support Mie-type resonances (*17-19*) at the target wavelength. Together, the periodicity and resonator response enable collective SLRs, which enhance scattering strength, promote bandgap formation, and improve optical feedback. Our simulations assumed periodic boundary conditions to model infinitely periodic metasurfaces and did not account for the finite size and perturbed periodicity of the metasurfaces. The optimized metasurface parameters are listed in table S1. Fig. 2B shows the simulated angle-resolved emission spectra of the designed metasurface averaged over both transverse electric (TE) and transverse magnetic (TM) polarizations of the incident light (*16*). We observe multiple bands corresponding to the diffractive coupling to the guided modes of the SU8 layer and lattice modes of the metasurface. For a metasurface with a period of 365 nm, the lasing high-Q mode, indicated by a white arrow, appears at a wavelength of 560 nm, which is spectrally adjacent to the emission peak of Rh6G and satisfies the second-order Bragg diffraction condition at a waveguide mode index of n$_{WG}$=1.53 (*8*). Importantly, we identify a stop-band between 560 nm and 565 nm (white and blue arrows in Fig. 2B) at the polar angle $\theta = 0$ (Γ point) with a gap-midgap ratio of $\frac{\Delta\omega}{\omega}$ of ~0.9 %, where ω and Δω represent the central angular frequency and the stop-band width, respectively. The stop-band width serves as a direct measure of the Bragg length, which corresponds to the number of lattice planes required for 1/e diffraction efficiency (*8,20-22*). In particular, the inverse of the relative stop-band width (0.9%) provides an estimate for the number of unit cells required for the Bragg diffraction (~100×100) in the periodic system. This approach allowed us to minimize the metasurface footprint while ensuring low lateral losses, thereby maintaining a high Q-factor essential for lasing. Fig. 2C shows the normalized electric field intensity distribution in the vertical y-z cross-section through the center of the unit cell for normal incidence at the resonant wavelength. Notably, the designed structure achieves a significant electric field intensity enhancement of over 2 orders of magnitude in the active medium.

Generating arbitrary emission patterns with metasurfaces typically requires aperiodic arrangements and/or in-plane variations in the shape and size of the constituent nanoresonators (*23-26*). To achieve this, in the second step, we employed inverse design using gradient-free global genetic optimization (*27*) to create an on-off binary metasurface by selectively removing or retaining nanoresonators. The optimization continued until the metasurface produced the desired far-field emission pattern "FSU", as the abbreviation of "Friedrich-Schiller-University".

The far-field pattern generated by the binary metasurface is determined by the convolution of the lasing mode angular emission profile with the Fourier transform (FT) of the real-space binary metasurface array (*23,24*). Here, the lasing mode is outcoupled via scattering from the nanoresonators, which act as secondary sources. The light from these secondary sources interferes, forming the far-field emission pattern, with their relative phases determined by the in-plane wave vector of the lasing mode ($k_\parallel = 0$) in our system. Assuming negligible distortions from metasurface truncation to 100×100 unit cells and periodicity perturbations, the secondary sources can be considered to all have the same scattering strength. Consequently, the far-field pattern is entirely governed by the spatial arrangement of the nanoresonators in the binary metasurface and the phase distribution imposed by the lasing mode, with the interference in the far-field representing the FT in the source plane (*24,28*). Further details of the optimization process are provided in (*16*) and fig. S2.

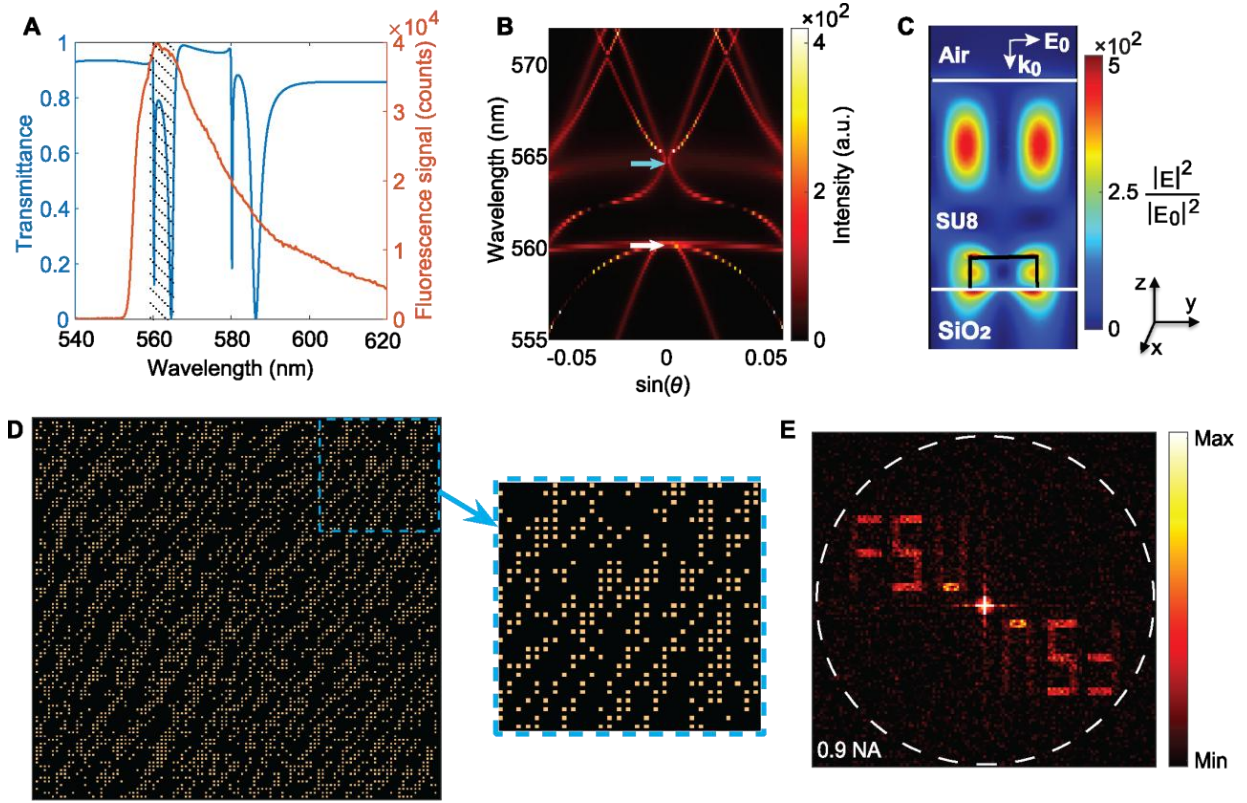

**Fig. 2. The two-step design process of on-off binary lasing metasurface.** First step: (**A**) Numerically calculated linear-optical transmission spectrum of the designed metasurface coated with SU8 layer for y-polarized normal incidence illumination (blue) and measured fluorescence spectrum of Rh6G (red) using a 550 nm band-pass filter. The range of interest is indicated with dashed hatching. (**B**) Simulated angle-resolved emission spectrum of the metasurface for azimuthal angle $\varphi = 0$ and polar angle $\theta$ up to ±3 degrees averaged over TE and TM polarization of the incident light. The white and blue arrows indicate the lower and upper edges of the stop-band, respectively. The color bar shows intensity enhancement with respect to the incident electric field intensity. (**C**) Calculated near-field intensity profiles at 560 nm in the vertical y-z cross-section through the center of the unit cell for normal incidence illumination, normalized with respect to the intensity of the incident plane wave. The outlines of the nanoresonator are shown in black. Second step: (**D**) The optimized arrangement of the nanoresonators of the binary metasurface for an output FT forming the letters "FSU". (**E**) The absolute value of the discrete FT of the optimized array in part (D).

Figure 2D shows the optimized spatial arrangement of the binary metasurface, where bright and dark pixels represent the presence or absence of nanoresonators, respectively. The metasurface measures approximately 37 μm × 37 μm, consisting of 100 × 100 unit cells (each 365 nm in size), with approximately 36% of the nanoresonators removed. Fig. 2E illustrates the absolute value of the discrete FT of the optimized array, which defines the far-field emission pattern of the binary lasing metasurface, revealing a central peak that corresponds to the initial periodic arrangement and two symmetrically arranged "FSU" in the second and fourth quarters. This symmetry arises from the inherent properties of FT.

We fabricated (*16*) a set of metasurfaces with varying design parameters to account for fabrication imperfections. The fabricated metasurfaces were spin-coated with SU8 photoresist mixed with 0.165 wt% Rh6G dye (*16*). A top-view scanning electron microscope (SEM) image of a typical fabricated binary metasurface and a focused-ion beam (FIB) cross-section SEM image of the metasurface after spin coating are depicted in Fig. 3A and its inset, respectively. The spin-coated SU8 film appears uniform and has a thickness of $h_{WG}$=880 nm. The dimensions of the fabricated metasurface, listed in table S1, closely match the optimized parameters from the design step.

Next, we experimentally characterized the fabricated metasurfaces. We pumped the sample using a 532 nm pulsed laser with a pulse duration of 0.5 ns and performed power-dependent fluorescence emission spectroscopy (see (*16*) and fig. S3). All measurements are performed with a laser pulse repetition rate of 1 Hz, with each acquisition integrating over a 1s period. Fig. 3B demonstrates the output emission intensity as a function of wavelength and pump pulse energy for the spin-coated metasurface. At a pump pulse energy of 7.9 nJ, we observe a broad spectrum featuring the spontaneous emission from Rh6G dye. Upon increasing the pump pulse energy to 7.95 nJ and beyond, we observe a narrow peak emerging at a wavelength of 560 nm, corresponding to the lasing mode in this system. The inset of Fig. 3B shows the total integrated output intensity (pink curve) and full-width at half-maximum (FWHM) (blue curve) as functions of input pulse energy. The output intensity exhibits the characteristic S-curve of lasing, marking transitions from spontaneous emission to amplified spontaneous emission, and finally to the lasing regime. Simultaneously, the FWHM narrows significantly above threshold, indicating the onset of spectral coherence.

Further, we performed back focal plane (BFP) imaging of emission (*16*) to investigate the evolution of the far-field radiation pattern of the metasurface from below to above the lasing threshold. All BFP images are measured using a 560 nm (10 nm bandwidth) band-pass filter in the detection path. Fig. 3C shows the BFP image below the threshold at 7.8 nJ in the spontaneous emission regime. The main features in this regime are the diffraction modes of the structure, which originate from the periodic arrangements of the nanoresonators. Note the convergence of separate diffractive modes in the normal direction (center of the BFP image), as characteristic of SLRs. As the pump pulse energy increases to 8 nJ, the system approaches the lasing threshold. In the corresponding BFP-image (Fig. 3D), this manifests as a weakening of the diffractive features of the spontaneous emission and the simultaneous emergence of a sharp peak at the center of the BFP image ($k_x$, $k_y$=0), where $k_x$ and $k_y$ are the in-plane components of the photon momentum. This highly directive peak indicates a coherent lasing emission and, more importantly, is accompanied by a subtle formation of our target pattern "FSU", already at this pump pulse energy. Finally, at a pump pulse energy well above the threshold (8.8 nJ) and after spatially filtering the strong central lasing peak for better visibility, we can observe (Fig. 3E) the target "FSU" pattern, generated by our binary lasing metasurface and showing a very good agreement with the simulated pattern in Fig. 2E.

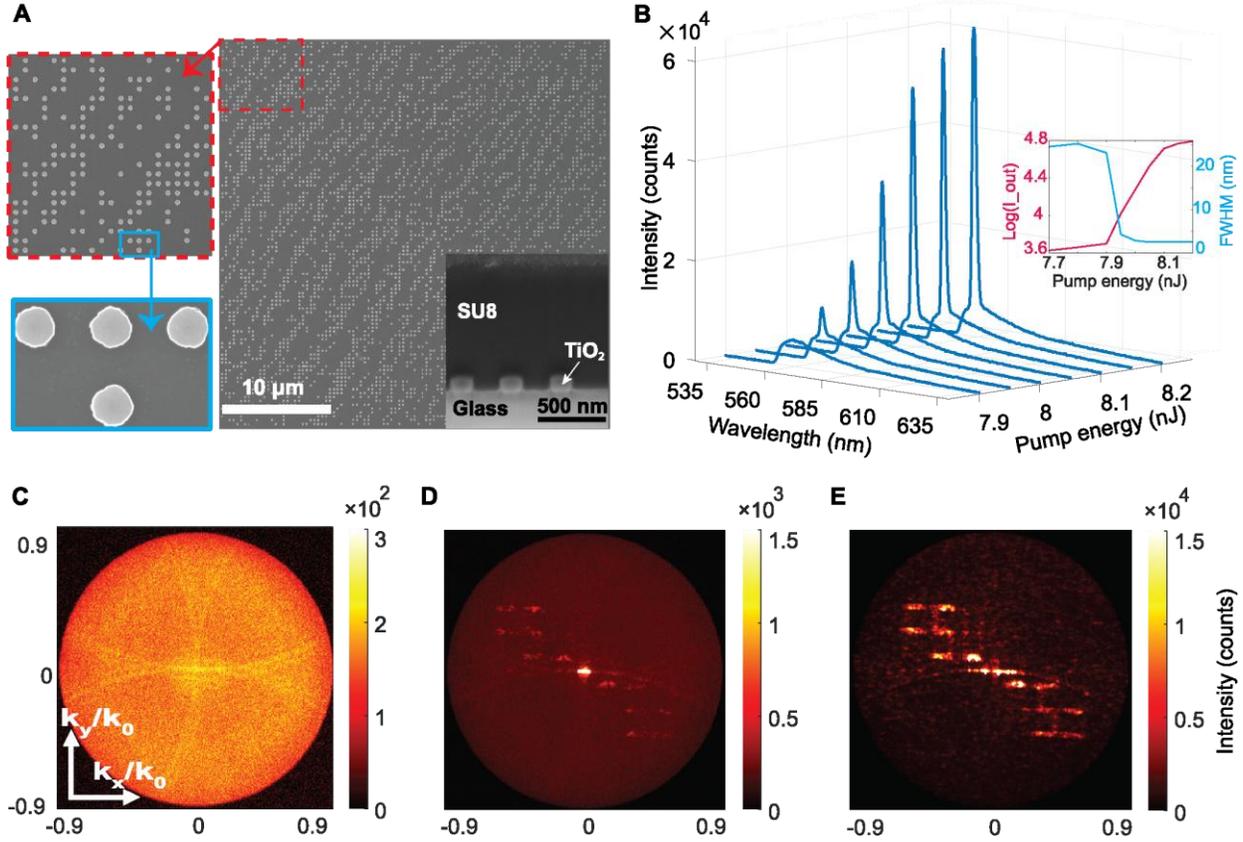

**Fig. 3. On-off binary lasing metasurface enabling coherent light generation with precise spatial control of the output beam.** (**A**) The top-view SEM image of a typical fabricated metasurface with a period of 365 nm. The inset shows a FIB cross-section of the metasurface coated with 880 nm SU8 film doped with Rh6G laser dyes. (**B**) Evolution of the emission spectra of the same metasurface at different pump pulse energies. The inset represents the maximum output intensity I_out (pink curve) and FWHM of the emission peak (blue curve) at different pump energies. (**C-E**) Measured BFP image of the lasing metasurface using a 560 nm band-pass filter: (**C**) Below threshold (7.8 nJ), it shows the diffraction bands of the spontaneous emission. (**D**) Right above threshold (8 nJ), the onset of lasing emission manifests as a bright, narrow feature in the center, accompanied by the appearance of the target "FSU" pattern. (**E**) Well past the threshold (8.8 nJ) and after spatially filtering the strong central peak, the target "FSU" pattern becomes prominent.

To further characterize the lasing emission, we performed momentum-resolved spectroscopy measurements by spectrally resolving a thin slice of the respective BFP images (*16*) around $k_y = 0$. Fig. 4A illustrates the results below the lasing threshold. Similar to the calculated angle-resolved emission spectra shown in Fig. 2B (covering a limited range of $\lambda$ and $\theta$ with fine resolution) and fig. S4 (spanning a broader range of $\lambda$ and $\theta$ with coarser steps for qualitative comparison), we observe highly dispersive narrow bands corresponding to the guided modes in the metasurface slab. However, due to the perturbed periodicity of the metasurface, limited resolution of our imaging spectrometer, and noise, fine features are not thoroughly resolved in the measured data. Nevertheless, we can identify the origin of the lasing mode at ($\theta = 0, \lambda = 560$ nm). Above threshold (Fig. 4B), the emission is concentrated at a single point of ($\theta = 0, \lambda = 560$ nm), demonstrating the high degree of spatial and temporal coherence of the lasing emission. The inset in Fig. 4B illustrates the $k_y = 0$ cross-section of the corresponding BFP image (Fig. 3D) with a green dashed line. Fig. 4C shows the same measurement repeated for the cross-section through the FSU pattern (inset of Fig. 4C). As we expected, the lasing emission appears as multiple dots constructing the top lines of the displayed letters, spectrally narrowed down to the same wavelength of 560 nm, but directed to the design angles. Thus, this binary metasurface shows great opportunities for spatially tailoring the self-generated lasing light.

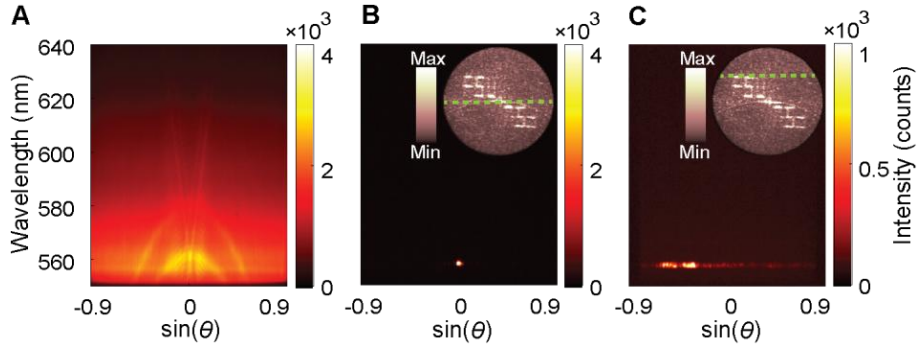

**Fig. 4. Momentum-resolved spectroscopy further approves full directional control of the self-generated lasing emission enabled by our on-off binary metasurface.** (**A**) A thin slice ($k_y = 0$) of the BFP image shown in Fig. 3C is spectrally resolved, showing the band diagram of the metasurface below the threshold. Note that for better visibility, the measurement below the threshold is performed with an excitation laser repetition rate of 4 kHz and integrated for 30 s. (**B**) Above the threshold, the emission is fully concentrated at the particular point of ($\theta = 0, \lambda = 560$ nm). Similar to (A), as shown in the inset with a dashed green line, the slice is cut from the center of the BFP image ($k_y = 0$). (**C**) Well above the threshold, where the target pattern emerges clearly, a cross-section through the upper part of the generated pattern (see inset) is spectrally resolved, revealing the lasing emission directed towards oblique angles at 560 nm.

Finally, we examined an additional fabricated metasurface (*16*, fig. S5), featuring the same spatial distribution of nanoresonators but slightly different nanoresonator dimensions, which supports multimode lasing action. The observed doubling of the target image (fig. S6, fig. S7) further underpins the rich engineering options offered by the combined effect of structure and array factor on far-field shaping.

We have analyzed and outlined the key design strategies and operational mechanisms enabling a unified platform that combines laser light generation and precise spatial manipulation within a flat and integrable device. Beyond its fundamental implications, our approach paves the way for the next generation of light-emitting devices, where tailored emission directionality can be harnessed for applications in optical communications, beam steering, LiDAR, optical neuromorphic computing, and high-resolution imaging.

**Acknowledgments:**

We thank B. Narantsatsralt, D. Arslan, and T. Bucher, for valuable discussions.

**Funding:** This work was funded by the Deutsche Forschungsgemeinschaft (DFG, German Research Foundation) through the International Research Training Group (IRTG) 2675 "Meta-ACTIVE", project number 437527638, and through the Emmy Noether Program, project number STA 1426/2-1.

**Author contributions:** I.S. conceived the idea. A.B. performed the design, simulations, the experimental setup building, and optical characterizations. A.V. contributed to the design and experimental setup preparation. M.Y. and K.T. performed the device fabrication. M.R. performed the FIB cut measurement. A.B. and I.S. contributed to the interpretation of results and participated in manuscript preparation.

**Competing interests:** The authors declare that they have no competing interests. **Data and materials availability:** All data needed to evaluate the conclusions in the paper are available in the manuscript or the supplementary materials.